\documentclass[reprint,showkeys,amsmath,amssymb,aps]{revtex4-2}
\usepackage{graphicx}
\usepackage{dcolumn}
\usepackage{bm}
\usepackage{xcolor}

\usepackage{hyperref}

\begin{document}  

\preprint{APS/123-QED} 

\title{Escape dynamics of a self-propelled nanorod from circular confinements with narrow openings}

\author{Praveen Kumar}
\affiliation{Department of Chemistry, Indian Institute of Technology Bombay, Mumbai 400076, India}
\author{Rajarshi Chakrabarti}
\email{rajarshi@chem.iitb.ac.in}
\affiliation{Department of Chemistry, Indian Institute of Technology Bombay, Mumbai 400076, India}


\begin{abstract}
\noindent We perform computer simulations to explore the escape dynamics of a self-propelled (active) nanorod from circular confinements with narrow opening(s). Our results clearly demonstrate how the persistent and directed motion of the nanorod helps it to escape. Such escape events are absent if the nanorod is passive. To quantify the escape dynamics, we compute the radial probability density function (RPDF) and mean first escape time (MFET) and show how the activity is responsible for the bimodality of RPDF, which is clearly absent if the nanorod is passive. The broadening of displacement distributions with activity has also been observed. The computed mean first escape time decreases with activity. In contrast, the fluctuations of the first escape times vary in a non-monotonic way. This results high values of the coefficient of variation and indicates the presence of multiple timescales in first escape time distributions and multimodality in uniformity index distributions. We hope our study will help in differentiating activity-driven escape dynamics from purely thermal passive diffusion in confinement.
\end{abstract}

\maketitle

\section{Introduction}\label{sec:Introduction}
\noindent \noindent The ``narrow escape problem" is an omnipresent phenomenon in areas like biophysics, polymer physics, cellular biology, \textit{etc.} \cite{schuss2007narrow,bressloff2013stochastic,holcman2015stochastic,grigoriev2002kinetics,co2017stochastic}. As the name suggests, this refers to a range of phenomena where a molecule or a nano or micron-sized object escapes from confinements through narrow opening(s). Examples from biology include ion transport through ion channels made of proteins, proteins trafficking through nuclear pore complex (NPC) or passing through mucus membrane, and bacteria motion through disordered media where it moves between the confined domains connected by narrow channels \cite{grigoriev2002kinetics,gorski2006road,misteli2008physiological,loverdo2008enhanced,chopra2022geometric,theeyancheri2022migration}. In particular, the movement of proteins between two chambers \textit{via} a narrow opening is an example of a narrow escape problem. In polymer physics, translocation is a type of phenomenon that involves the transport of macromolecules through a narrow opening and is, therefore, closely related to the narrow escape problem \cite{muthukumar1999polymer,sebastian2000kramers,chen2021dynamics}. One way to quantify the narrow escape problem is to compute the mean first-passage time, which is the average time to exit through the opening \cite{benichou2008narrow,grebenkov2017diffusive,srivastava2021brownian,grigoriev2002kinetics}. In recent years, narrow escape problems have received increasing attention from researchers due to their relevance and potential applications in diverse fields, such as cellular biophysics, site-specific delivery of drugs, and nonviral gene delivery \cite{behzadi2017cellular,mitchell2021engineering,schuss2007narrow,bressloff2013stochastic,holcman2015stochastic}. \\

\noindent The escape dynamics of passive Brownian walkers from different types of confined geometries have been extensively studied, both experimentally and theoretically \cite{holcman2015stochastic,burada2009diffusion,bosi2012analytical,hanggi2009artificial,kullman2002transport,biess2007diffusion,srivastava2021brownian,agranov2018narrow,grebenkov2019full,ghosh2011geometric,muthukumar1999polymer,sebastian2000kramers,grebenkov2017diffusive}. Therein, the escape dynamics of the passive particles has been analyzed by varying the control parameters, such as the shape and volume of confined geometries, the size of the pore and the randomly moving particle, and the number of openings (small pores) \cite{grebenkov2017diffusive,agranov2018narrow,srivastava2021brownian,kullman2002transport,hanggi2009artificial,chevalier2010first}. Thus, the escape dynamics of passive Brownian particles from confined regions represent a complex interplay among these factors, the properties of the surrounding environments, and the nature of the escaping particles \cite{bressloff2007diffusion,biess2007diffusion,bressloff2013stochastic,gorski2006road,misteli2008physiological,holcman2004escape,freche2011synapse,holcman2015stochastic,chen2022first}. \\

\noindent The diffusion of self-propelled objects in complex environments is a fascinating and promising area of research that still offers ample opportunities for further exploration and investigation \cite{bechinger2016active,theeyancheri2022silico,ligesh2023,kumar2023dynamics,narinder2019active,caprini2018active,jakuszeit2019diffusion,chepizhko2013diffusion,ghosh2021active,caprini2020diffusion}. It becomes even more interesting when self-propelled objects navigate through confined geometries with narrow openings \cite{nayak2020escape,debnath2021escape}. Here, the term self-propelled objects refers to biological or artificial agents that consume energy from the environment or utilize their internal chemical energy to get directed motion in a complex environment \cite{ramaswamy2010mechanics,mujtaba2021micro}. Examples include biological entities such as bacteria \cite{zhang2010collective,sokolov2007concentration}, spermatozoa \cite{fauci2006biofluidmechanics}, insects, and animals \cite{buhl2006disorder,tunstrom2013collective} and synthetic systems such as self-propelled Janus rods, discs, light-induced active particles, and chemically driven systems \cite{kudrolli2008swarming,jiang2010active,pourrahimi2018multifunctional,walther2013janus,ignes2022experiments,palacci2013living,kapral2013perspective,mujtaba2021micro}. Furthermore, because of the additional energy sources, these objects are typical non-equilibrium systems with peculiar transport and collective properties \cite{howse2007self,buttinoni2013dynamical,zhang2010collective,fily2012athermal,mujtaba2021micro}.\\

\noindent In recent years, some theoretical and experimental studies have been reported on the quantification of the mean first passage time of the self-propelled particles in confined geometries, such as a Petri dish, a pie-wedge shape, and various shaped confinements with narrow openings  \cite{biswas2020first,debnath2021escape,ghosh2014communication,caprini2019transport,khatami2016active,olsen2020escape,paoluzzi2020narrow,biswas2023}. A very recent experimental study reported that the mean first passage times of an active particle rise monotonically as a function of the area fraction of the surrounded passive discs (crowders), while their fluctuations exhibit non-monotonic behavior \cite{biswas2020first}. Debnath \textit{et al.} numerically investigated the mean exit time of an active particle from circular confinement with single or multiple exit windows, and they observed that overdamped active particles spread on the surface of the confinement at extensive self-propulsion lengths and rotational dynamics governs the escape process \cite{debnath2021escape}. A recent numerical study investigated the effect of motility parameters on the narrow escape time of active particles from circular domains \cite{paoluzzi2020narrow}. This study compared two paradigmatic models of active motions, namely, run-and-tumble and active Brownian. It has been found that run-and-rumble particles are less efficient in escaping from the chamber compared to active Brownian particles at equal persistence times in the active limit. The escape dynamics from a circular disk for interacting active particles has further been extended using a minimal two-dimensional Vicsek model \cite{olsen2020escape}. The results of this study revealed that the survival probability for noninteracting particles is exponential, whereas the interacting particles have an exciting crossover of survival probability from initial exponential to subexponential late-time decay. In a very recent study, it has been shown how worms escape from one confined domain to another through a narrow channel \cite{biswas2023}. Mostly, previous attempts focused on the confinement shape and relative size of the active particle to the small pore and the effect of crowding to study the escape dynamics. However, much less is explored on the effect of the shape of particles on their dynamics inside bounded domains with a single and multiple narrow opening(s). Anisotropy in shape leads to more complicated dynamical behaviors. Additionally, the presence of multiple targets in the form of small openings complicates the escape dynamics from the confinement. The role of confinement on the dynamics of active particles is unquestionably critical in realistic systems, particularly for biological matter and biotechnology. Even though there has been some progress to date, a lot of fundamental questions still need to be resolved. For instance, how the self-propulsion and the number of openings located on the boundary affect the escape dynamics of the nanorod from the circular domain. The fields of ``active transport" and ``narrow escape" are currently of great interest, and this work serves as an example that integrates both of these research topics.\\

\begin{figure*}[t]
    \centering
    \includegraphics[width=0.92\linewidth]{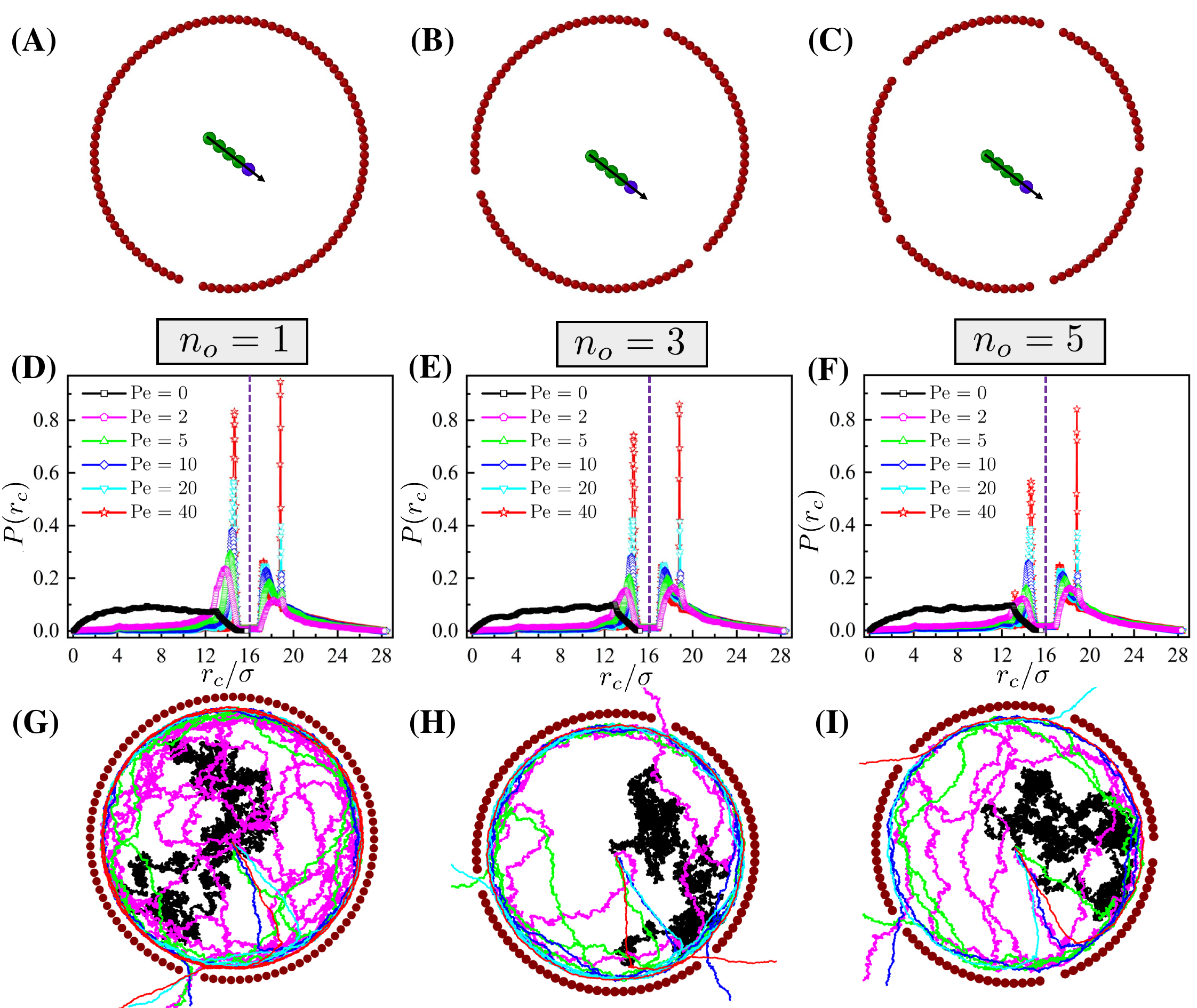}
    \caption{The snapshots of a self-driven nanorod inside a circular confinements with (A) single, (B) three, and (C) five narrow opening(s) through which the nanorod escapes or enters the confinement. The self-propulsion is directed from green to blue beads, and the arrow represents it in (A), (B), and (C). The radial probability density distributions ($P(r_c)$) of the nanorod for the confinement with single, three, and five opening(s) at different activities (Pe) are shown in (D), (E), and (F), respectively. The corresponding COM trajectories of the nanorod in (D), (E), and (F) are shown in (G), (H), and (I), respectively, for different Pe. The dashed purple vertical lines at $r_c = 16\sigma$ in (D), (E), and (F) represent the boundary of the domain, where ${r_c}<16\sigma$ and ${r_c}>16\sigma$ are the inside and outside of the confinement, respectively.}
    \label{fig:Model_RDF_Traj}
\end{figure*}

\noindent In this paper, we study the narrow escape dynamics of a self-propelled rigid nanorod confined to an impenetrable circular domains with single and multiple narrow openings (small open windows) through which the nanorod may escape [Fig.~\ref{fig:Model_RDF_Traj}(A-C)]. We perform Brownian dynamics simulations to unveil the escape dynamics of a self-propelled nanorod through the narrow opening(s). For this purpose, we analyze the dynamics and statistics of the first escape times for the nanorod. We compute the radial probability density function by following the centre of mass (COM) position $r_c$ of the rigid nanorod. We found that a passive nanorod shows restricted movement inside the circular confinement, whereas the self-propelled nanorod moves faster, reaches the boundary, and escapes through the narrow opening(s) present at the boundary. This escape is assisted by the persistent and directed motion of the nanorod. Our analyses of the self-part of the van-Hove correlation function of the nanorod show that the correlations become flat at higher displacement and broader with shoulder peaks by increasing the self-propulsion. In addition, the mean first escape (exit) times of the self-propelled nanorod monotonically decrease by increasing the activity and the number of narrow openings on the boundary of the confinement. At the same time, their fluctuations (coefficient of variation $CV$) show non-monotonic behavior. To understand the fluctuations in trajectories (first escape times), we look at the full distribution $F(\tau_{es})$ and uniformity index distribution $P(\omega)$ of random first escape times of the active nanorod. We believe that our investigations contribute to understanding the escape dynamics of a biomimetic object of elongated shape, a nanorod in this case, from confined geometries with small open windows. \\ 

\noindent This paper is organized as follows. In Section \ref{sec:Model}, we explain the model system investigated in the present work. First, we describe the method for computationally constructing the self-driven rigid nanorod, which diffuses inside the circular rigid confinement and escapes through a small window. Thereafter, we describe the simulation protocol employed in our simulation study. Results and discussion are presented in Section \ref{sec:Results}. Finally, we conclude the paper in Section \ref{sec:Conclusion}. \\

\section{Model and simulation details}\label{sec:Model}
\begin{figure}[t]
    \centering
    \includegraphics[width=0.9\linewidth]{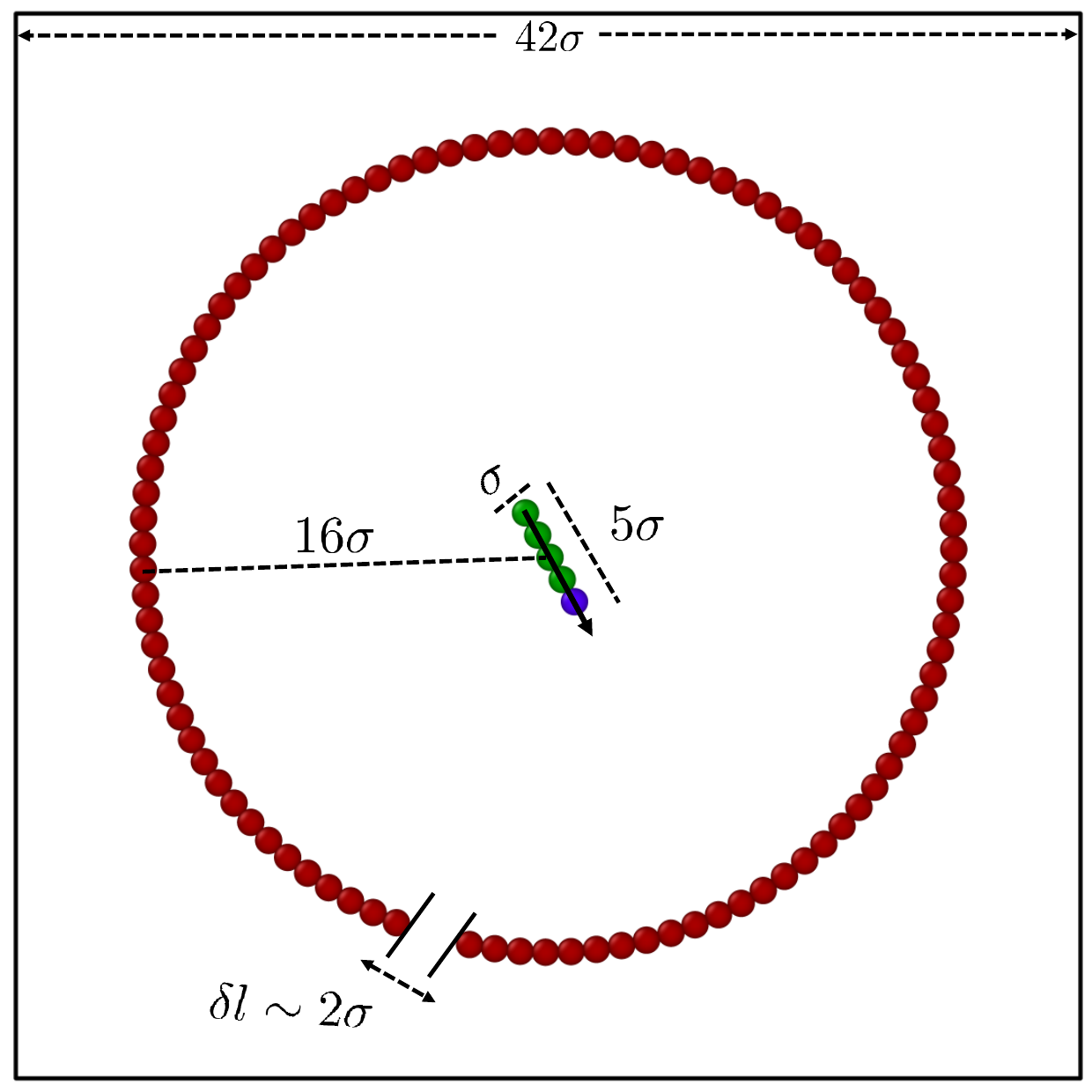}
    \caption{The snapshot of a self-driven nanorod of length $5\sigma$ inside a circular confinement with a single narrow opening. The size of the narrow opening is around $\delta l = 2\sigma$, which is created by removing the pair of particles from the circumference of the ring confinement. The black arrow indicates the direction of self-propulsion. Ovito package is used to generate this snapshot \cite{stukowski2009visualization}.}
    \label{fig:model}
\end{figure}

\noindent We model the microswimmer as a self-driven rigid nanorod which is composed of five beads, each with an identical diameter $\sigma$ and mass $m$, where $\sigma$ has the dimension of length. The beads of the nanorod are connected by harmonic potential, denoted by $V_\text{harmonic}$, and expressed as follows:

\begin{equation}
V_{\text{harmonic}}\left(r_{ij} \right)= k_{\text{H}} \frac{(r_{ij} - r_{0})^2}{2}
\label{eq:harm4}
\end{equation}

\noindent here $k_{\text{H}}$ is the spring constant and $r_{0}$ represents the equilibrium distance between two neighboring beads of the nanorod, which is constant throughout the simulations. Thus, the nanorod has a fixed length of $5\sigma$. We ensure that the nanorod always moves and rotates as a single rigid body. Next, circular confinement with different numbers of narrow openings is modeled as rigid circular confinement with a radius of $16\sigma$, and the boundary (circumference) of the confinement is made of circular beads of diameter $\sigma$ [see Fig.~\ref{fig:model} for the confinement with a single narrow opening]. Each of the narrow openings is created by removing a pair of particles from the boundary of the confinement so that the size of each narrow opening (small window) is fixed ($\delta l \sim 2\sigma$), which is much smaller than the size of circular rigid confinement and also much smaller than the length of the nanorod. This ensures that the nanorod can only escape through these holes/openings head-on. The circular domain is enclosed in a two-dimensional square box with edge length $42\sigma$. The center of the ring confinement is fixed at the coordinates (0,0) of the box during the simulations. The rigid nanorod with random orientation is carefully placed at the center of the circular rigid confinement with the narrow opening(s) through which it can exit or enter. To accurately capture the entire process of escape and reentry of a self-driven rigid nanorod through a narrow opening, we have set the length of the simulation box to be ten units larger than the size of the circular confinement. In this work, we systematically investigated three cases by changing the number of small openings in the system, specifically by examining cases with a single opening, three openings, and five openings. These openings, referred to as small pores or windows, are intentionally positioned in a well-separated manner. Here, we aim to gain insights into how the number of openings influences the escape dynamics of the nanorod from the circular confinement \cite{lagache2017extended,chevalier2010first}. Confinements or domains with multiple small openings can be seen as a scenario where search processes occur for multiple targets. Mostly, previous models related to narrow escape problems (first passage time problems) focused on a single target (exit) in a given search domain \cite{paoluzzi2020narrow,condamin2005first,benichou2008narrow,agranov2018narrow,grebenkov2017diffusive}. \\ 

\noindent The non-bonded pairwise interactions between all the particles of the confinement and the nanorod are set as purely repulsive and modeled by Weeks–Chandler–Andersen (WCA) potential  \cite{weeks1971role}: 
\begin{equation}
V_{\text{WCA}}(r_{ij})=\begin{cases}4\epsilon_{ij}\left[\left(\frac{\sigma_{ij}}{r_{ij}}\right)^{12}-\left(\frac{\sigma_{ij}}{r_{ij}}\right)^{6}\right]+\epsilon_{ij}, \hspace{1mm} \mbox{if } r_{ij} < r_{cut}\\
=0, \hspace{40mm} \mbox{otherwise}
\end{cases}
\label{eq:WCA4}
\end{equation}
\noindent where $r_{cut} = 2^{1/6}\sigma_{ij}$, $\sigma_{ij} = \frac{\sigma_i + \sigma_j}{2}$, with $\sigma_{i(j)}$ being the diameter of the interacting pairs. $\epsilon_{ij} = 1$ and $r_{ij}$ are the strength of the steric repulsion and the distance between the interacting particles, respectively. \\

\noindent The following Langevin equation is implemented to describe the dynamics of each particle with mass $m$ and position $r_i(t)$ at time $t$,

\begin{equation}
m\frac{d^2 \textbf{r}_{i}(t)}{dt^2} = - \gamma \frac{d \textbf{r}_{i}}{dt} - \sum_{j} \nabla V(\textbf{r}_i-\textbf{r}_j) + {\bf f}_{i}(t) + {\bf{F}_{\text{a}} \bf \hat{n}} 
	\label{eq:Langevineq4}
\end{equation}

\noindent where $\gamma$ is the friction coefficient, which is very high ($5.3\times 10^{4}$) in our simulations so that the underlying dynamics is effectively overdamped. $V(r) = V_{\text{harmonic}} + V_{\text{WCA}}$ is the resultant interaction potential of the system [Eq.~\ref{eq:harm4} and Eq.~\ref{eq:WCA4}]. Thermal fluctuations are captured by the Gaussian random force ${\bf f}_{i}(t)$, satisfying the fluctuation-dissipation theorem.
\begin{equation}
	\left<f(t)\right>=0, \hspace{5mm}
	\left<f_{\alpha}(t^{\prime})f_{\beta}(t^{\prime\prime})\right>=4k_B T \gamma  \delta_{\alpha\beta}\delta(t^{\prime}-t^{\prime\prime})
	\label{eq:random-force_4}
	\end{equation}
\noindent where $k_{B}$ is the Boltzmann constant, $T$ is the temperature, and $ \delta$ represents the Dirac delta function. $\alpha$ and $\beta$ are the Cartesian components. The activity is introduced as a propulsive force $\bf{F}_{\text{a}} \bf \hat{n}$, where $\text{F}_\text{a}$ represents the magnitude of the constant self-propulsion with orientation specified by the unit vector $\bf \hat{n}$ along the body axis of the nanorod. The inclusion of active force breaks the fluctuation-dissipation theorem. In the present study, the Lennard-Jones parameters $\sigma$, $\epsilon$, and mass $m$ are the fundamental units of length, energy, and mass, respectively. All other physical quantities are therefore reduced accordingly and expressed in terms of the fundamental units $\sigma$, $\epsilon$, and mass $m$, and presented in dimensionless forms. The energy and time units have been designated as $k_B T$ and $\sqrt{\dfrac{m\sigma^2}{k_{B}T}}$, respectively. To measure the strength of active force (self-propulsion), we express it in terms of a dimensionless quantity Pèclet number Pe, defined as $\frac{\text{F}_\text{a} \sigma}{k_B T}$. Therefore, Pe = 0 corresponds to the passive nanorod. \\

\noindent All simulations are performed in a square box with periodic boundary conditions in all directions. We use Langevin thermostat by employing the LAMMPS package \cite{plimpton1995fast}, and the equation of motion [Eq.~\ref{eq:Langevineq4}] is integrated using the velocity Verlet algorithm in each time step. All the production simulations are carried out for $5 \times 10^8$ steps, and the simulation integration time step is set to be $5 \times 10^{-4}$. We record the positions of the particles for every $100th$ step. \\ 

\noindent We have carried out 150 independent simulations for a given value of Pe and the number of openings, $n_o$. The Pe values considered here are 0, 2, 5, 10, 20, and 40, where Pe = 0 stands for the passive case. We consider three values of $n_o$, \textit{viz.} $n_o = 1$ (a single opening), $n_o = 3$ (three openings), and $n_o = 5$ (five openings). For each case, every simulation starts therefore placing the nanorod at the centre of the confinement, and they run independently. \\

\section{Results and discussion}\label{sec:Results}
\subsection{Dynamics and interaction of the nanorod with the boundary of the confinement}

\begin{figure*}[t]
    \centering
    \includegraphics[width=0.95\linewidth]{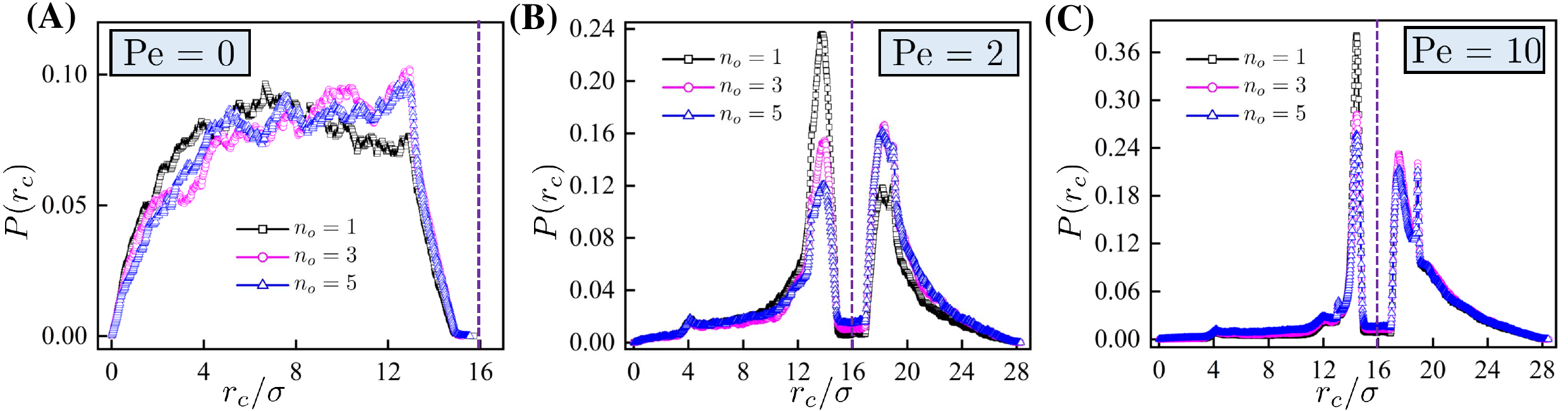}
    \caption{$P(r_c)$ of the self-propelled nanorod at (A) $\text{Pe} = 0$, (B) Pe $ = 2$, and (C) $\text{Pe} = 10$ for confinements with single, three, and five opening(s). The dashed purple vertical lines at $r_c = 16\sigma$ represent the boundary of the circular domain.}
    \label{fig:RDF_comp}
\end{figure*}

\noindent In order to understand the escape dynamics of a self-driven nanorod from the circular confinement through narrow opening(s), we first compute the radial probability density function (RPDF) $P(r_c)$ of finding the nanorod at different active forces (Pe) for all three cases ($n_o = 1, 3, 5$), as shown in Fig.\ref{fig:Model_RDF_Traj}(A-C). Here, $r_c$ is the centre of mass (COM) position of the nanorod. In this study, the radial probability density functions are normalized so the total area under each RPDF curve equals unity. From Fig.~\ref{fig:Model_RDF_Traj}(D-F), we notice that the passive (Pe = 0) nanorod has almost equal probability all over the place inside ($r_c < 16 \sigma$) the circular confinement but drops at the boundary. This indicates the passive nanorod diffuses smoothly inside the confinement with no particular preference to the boundary of the circular confinement [see ESI videos, Movies S1-S3 for $n_o = 1, 3, $ and $5$, respectively]. As mentioned, $P(r_c)$ drops quite sharply close to the boundary, which accounts for the repulsive interaction between the nanorod and the boundary particles. This observation ensures that the passive nanorod since has no attractive interaction with the boundary, it spends most of its time away from the boundary and therefore never escapes through the opening(s) present at the boundary. At least, we do not capture any escape events for the passive nanorod in our simulations. In all the cases with $\text{Pe} > 0,$ RPDFs are bimodal and these bimodal distributions have two peaks with maxima, one inside and the other outside the circular confinement [Fig.~\ref{fig:Model_RDF_Traj}(D-F)]. This indicates that the active nanorod escapes through the narrow opening(s). The peaks inside the confinement appear around $r_c \approx 14\sigma$ (the boundary is at $16\sigma$), indicating the fact that the active nanorod has spent most of its time close to the boundary. In the case of an active nanorod, there are events where the active nanorod goes to the boundary and persistently pushes the boundary before rotating and moving along the boundary. This can also be seen clearly in ESI videos Movies S7 and S15. This is a characteristic of active Brownian particle (ABP), a result of its persistent motion. Escape events of active nanorods can also be clearly seen from ESI videos, namely, Movies S4, S5, and S6 for all three cases. This requires a proper alignment of the nanorod in proximity to the boundary. To support this understanding, we have taken some snapshots from the simulation of the nanorod successfully escaping through various narrow openings present on the confinement's boundary. These snapshots depict different orientations of the nanorod with respect to the openings during the escape process [Fig.~S2, ESI]. Conversely, we have also taken snapshots where the nanorod is in close proximity to the hole but still unable to escape due to its improper orientation, as it appears to be nearly parallel to the wall [Fig.~S3, ESI]. On the other hand, the second peak at $r_c \approx 19\sigma$ represents the probability of finding the active nanorod outside the confinement when it is sticking to the impenetrable boundary. But this sticking is due to its persistent motion when aligned roughly perpendicular to the boundary wall from outside. This happens without any sticky LJ (Lennard-Jones) interaction of the nanorod with the boundary. This mode of approach actually helps the nanorod to re-enter through the opening(s). We manifest that the nanorod prefers to stay outside, close to the boundary of the confinement, until it gets a chance to re-enter the circular domain. In addition, a second sharp peak outside the confinement starts appearing at $r_c \approx 17\sigma$ with increasing Pe. It would be another possible approach for the nanorod to the boundary wall, where it is sliding against the wall rather than sticking to it. Both of these are captured in $P(r_c)$ peaks outside the confinement.\\

\noindent A careful observation shows that these apparently bimodal distributions have finer structures. The broad peak inside the confinement, on increasing Pe, becomes sharper and shifts close to the boundary. The same happens with the peak outside the confinement, it sharpens and shifts close to the boundary. This may be seen in the ESI videos, namely Movies S7, S8, and S9, at a higher Pe = 20 for all three cases. We also note that the probability of finding the active nanorod decreases inside while increasing outside the confinement at a higher activity (Pe = 40), as the heights of the peaks reflect in $P(r_c)$ [Fig.~\ref{fig:Model_RDF_Traj}(D-F)]. Similar behavior is observed for all three cases, namely single, three, and five narrow opening(s). Furthermore, the radial probability distributions are supported by the corresponding plots of the COM trajectory of the nanorod, as shown in Fig.~\ref{fig:Model_RDF_Traj}(G-I). From the trajectories, we can see clearly that the nanorod has uniform spreading inside the confinement in the passive case, while the active nanorod has circular spreading along the boundary. \\

\begin{figure*}[h]
    \centering
    \includegraphics[width=0.99\linewidth]{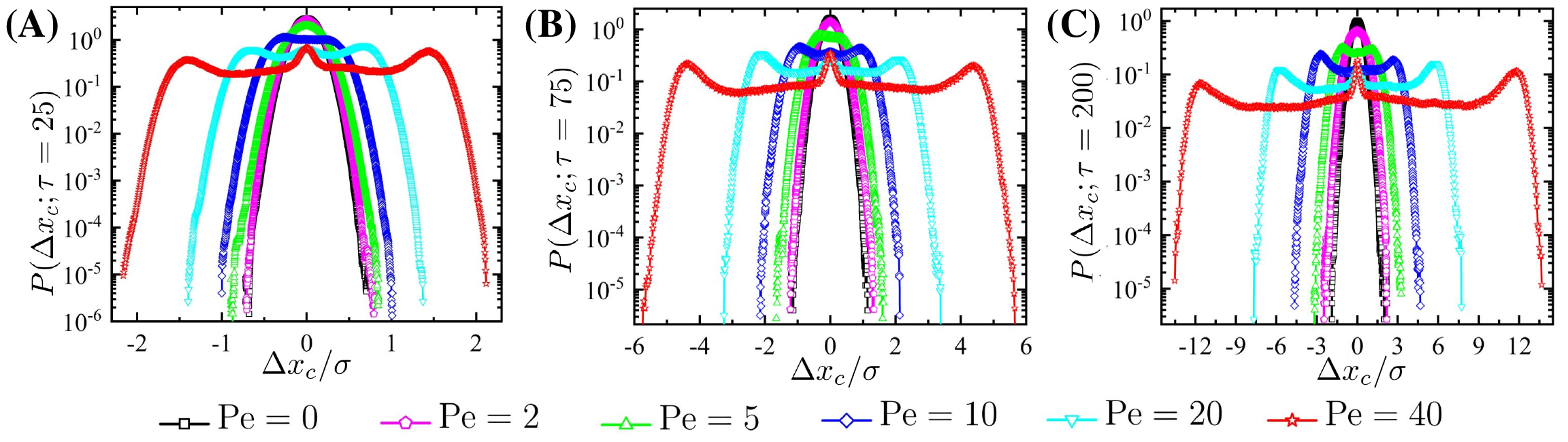}
    \caption{Plots (A)-(C) represent displacement distribution functions $P(\Delta x_{c}; \tau)$ of a nanorod at different lag times $\tau = 25, 75, 200$ for the confinement with three narrow openings.}
    \label{fig:PDF_3_SW_dX}
\end{figure*}

\noindent In Fig.~\ref{fig:RDF_comp}(A-C), we compare the radial probability distributions by changing the number of narrow openings while keeping Pe constant. By increasing the number of narrow openings, the number of events having escape and re-entrance increases, leading to a decrease in the probability of finding the active nanorod near the boundary of the confinement. The presence of multiple openings in confinement increases the accessibility and options for the active nanorod to find an escape route. This results in an increased probability of successful escape and re-entry for the nanorod, enhancing its ability to overcome confinement. This can also be clearly seen in ESI videos, namely Movies S5, S6, S8, and S9. Self-propulsion, as well as the number of narrow openings, play a crucial role in assisting the escape of nanorods from confinement. The presence of extended long tails away from the circular wall in the profile of $P(r_c)$ provide insights into the position probability distribution of the active nanorod both inside and outside of the confinement [Fig.~\ref{fig:Model_RDF_Traj}(D-F) and Fig.~\ref{fig:RDF_comp}]. This behavior leads to a low probability of finding the nanorod away from the circular wall. Additionally, we explored a scenario without any narrow openings on the confinement's boundary, and remarkably, the extended tails outside the circular wall in the profile of $P(r_c)$ disappear [Fig.~S4, ESI] \cite{narinder2019active}. These long tails outside the wall are a consequence of the narrow opening(s) present on the confinement's boundary. \\

\noindent In addition, to have an even deeper understanding of the nanorod motion, we analyze the self part of the van-Hove correlation function (the displacement probability distribution function) defined as $P(\Delta x_c; \tau) \equiv \left\langle \delta(\Delta x - (x_c(t+\tau)-x_c(t)))\right\rangle$ of the nanorod in one dimension, where $x_c(t+\tau)$ and $x_c(t)$ are the COM positions of the nanorod along $x$-direction at time $(t+\tau)$ and $t$, respectively. We compute the van-Hove self-correlation functions for a given time interval $\tau$ and study the effect of self-propulsion on $P(\Delta x_c)$ by varying Pe [Fig.~\ref{fig:PDF_3_SW_dX}]. The distribution curves for a passive (Pe $= 0$) and weakly active (Pe $= 2$) nanorod overlap, but the nanorod with high self-propelled velocities tends to have even broader and flatter distributions. This is consistent with the larger displacement of the active nanorod, both inside and outside the confinement [Fig.~\ref{fig:PDF_3_SW_dX}(A-C)]. More interestingly, there exist shoulder peaks in $P(\Delta x_c; \tau)$ at higher values of Pe. We also observed the shoulder peaks for moderate activities (say, Pe = 5, 10), but at larger $\tau$ [Fig.~\ref{fig:PDF_3_SW_dX}C]. This is a consequence of the impenetrable circular boundary, where the active nanorod stays for a longer time due to its directed motion toward the boundary. Typically, the particle with high self-propulsion tends to move towards the boundary of the confinement and escapes if it gets a chance \cite{bechinger2016active,kumar2022chemically,narinder2019active}. Similar distributions of displacements are observed earlier in an experimental study of an active particle confined to a glass Petri dish \cite{biswas2020first}. Furthermore, the flattening of the van-Hove distribution indicates that at higher self-propulsion, the nanorod moves freely inside and outside the confinement. In other words, shorter and larger displacements $\Delta x_c$ become equally probable. This behavior indicates that the nanorod has increased mobility and escape capability at higher Pe. Moreover, we also depict trial fittings of these van-Hove distributions with a Gaussian distribution, $P_g (\Delta x_c; \tau)=\frac{1}{\sqrt{2\pi\left<\Delta x_{c}^2\right>}}\exp\left({-\frac{\Delta x_{c}^2}{2\left<\Delta x_{c}^2\right>}}\right)$. Therefore, any deviation of the van-Hove distributions from the Gaussian curves indicates non-Gaussianity [Fig.~S5, ESI]. We found that the distributions are Gaussian for a given $\tau$ in the case of passive nanorod, whereas $P(\Delta x_c)$ starts deviating from the Gaussianity with increasing Pe owing to the self-driven motion of the nanorod. The deviation from Gaussianity becomes more pronounced at larger values of $\tau$ and Pe. These findings contribute to a better understanding of the effect of self-propulsion on nanorods in confined domains and their subsequent escape.

\subsection{Analyses of escape time}
\noindent Utilizing our existing knowledge of dynamics and the interaction of the nanorod with the boundary, we next discuss the results for the statistics of first escape times and specifically highlight the non-monotonic variation of first escape times by varying the magnitude of the active force and the number of narrow openings. We compute the mean first escape time (MFET), defined as the time taken by the particle to reach the boundary of the domain and escape through a narrow opening, as defined and used in the literature \cite{benichou2008narrow,biswas2020first,grebenkov2019full,mattos2014trajectory,chen2022first,mejia2011first,chevalier2010first,condamin2005first,mattos2012first,chakrabarti2009lower,biswas2023}. This can also be termed as ``mean first passage time (MFPT)". MFET in this context is essentially the MFPT. Furthermore, from the set of first escape times $\{\tau_{es}\}$, we directly compute the MFET $ \left < \tau_{es} \right >$ which is defined as follows,

\begin{equation}
    \left < \tau_{es} \right > = \int_{0}^{\infty} \tau_{es} F(\tau_{es}) d\tau_{es}
\end{equation}

\begin{figure*}[ht]
    \centering
    \includegraphics[width=0.99\linewidth]{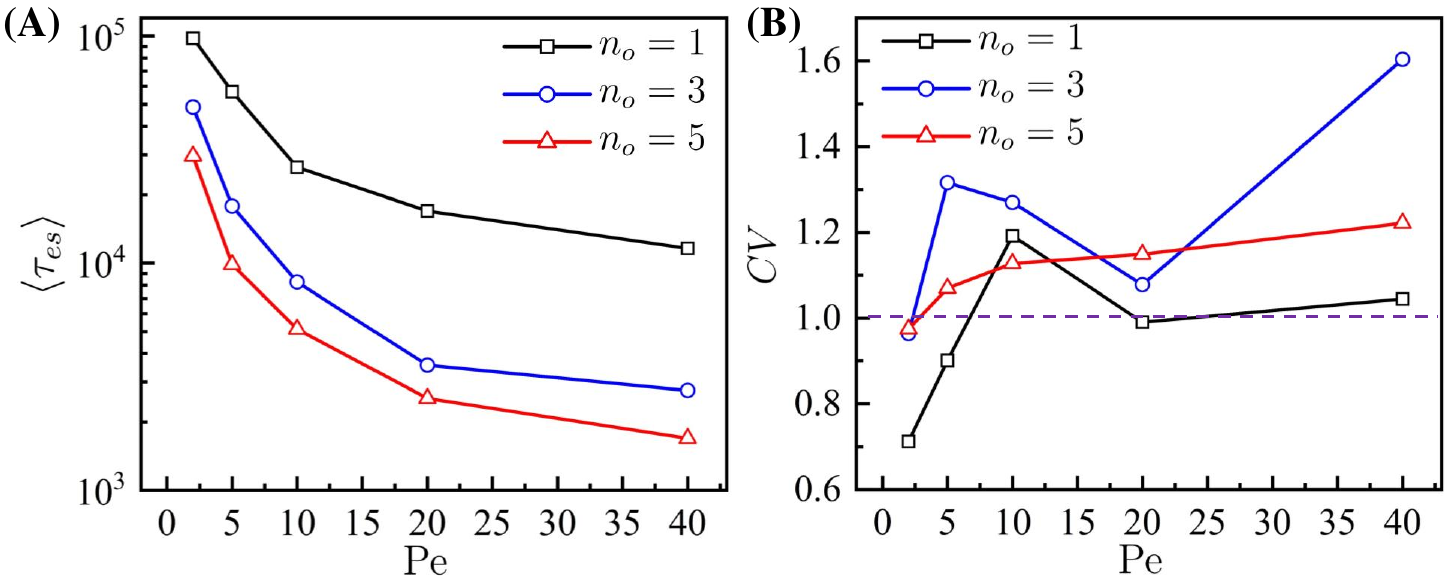}
    \caption{Plot of (A) mean first escape time (MFET) $\left < \tau_{es} \right >$ and (B) coefficient of variation ($CV$) \textit{versus} Pe. Here, one can observe the monotonic decrease of $\left < \tau_{es} \right >$ with an increase in Pe and the non-monotonic change of $CV$ as a function of Pe.}
    \label{fig:MFPT_CV}
\end{figure*}

\noindent where $F(\tau_{es})$ is the first escape time distribution. In our simulations, to detect first escape events we continuously track the center of mass position of the nanorod, considering an escape successful when the COM position of the nanorod is outside the confinement's boundary. We count the total number of steps taken by the nanorod to achieve the first escape and then multiply it by the simulation time. By doing so, we obtained the first escape time and by averaging all the first escape times, we computed the mean first escape time, which provides an insightful measure of the nanorod's escape behavior in the simulations. \\

\noindent In Fig.~\ref{fig:MFPT_CV}(A), we plot $\left < \tau_{es} \right >$ $versus$ Pe for all three cases, confinement with single, three, and five narrow openings ($n_o =$ 1, 3, and 5). One can observe that $\left < \tau_{es} \right >$ decreases monotonically with Pe as it is varied from 2 to 40 \cite{caprini2021correlated,chaki2020escape,goswami2023effects}. Self-propulsion always causes faster dynamics with increasing Pe, and as a result, the nanorod takes, on average, less time to reach the boundary and successfully escapes through the small open window. Higher values of Pe imply that the nanorod's active forces are stronger compared to the random motion resulting from diffusion, as a consequence, the nanorod performs persistent motion (directed motion). We also noticed that $\left < \tau_{es} \right >$ decreases monotonically with an increase in the number of openings on the impermeable confinement [Fig.~\ref{fig:MFPT_CV}(A)]. The confinement having multiple openings offers the nanorod different escape routes. If the nanorod is unable to escape through one of the exits, there is a finite probability that it will escape through the other opening(s). This multiplicity of exits significantly enhances the probability of finding a viable pathway for escape, particularly in complex or crowded environments. Furthermore, we plot $\dfrac{\left < \tau_{es} \right >}{\left < \tau_{es} \right >_{Pe = 2}}$ as a function of Pe to evaluate the extent of the reduction in mean first escape time caused by activity [Fig.~S6, ESI]. By comparing the values with a reference case at a specific Pe value (Pe = 2) for all the cases ($n_{o} = 1, 3, 5$), we observed a sharp decline in $\left < \tau_{es} \right >$ as Pe increases from 2 to 5 and from 5 to 10. However, beyond Pe of 10, the reduction in $\left < \tau_{es} \right >$ shows a relatively less pronounced difference as Pe continues to rise up to 40 [Fig.~S6, ESI].\\

\begin{figure*}[ht]
    \centering
    \includegraphics[width=0.99\linewidth]{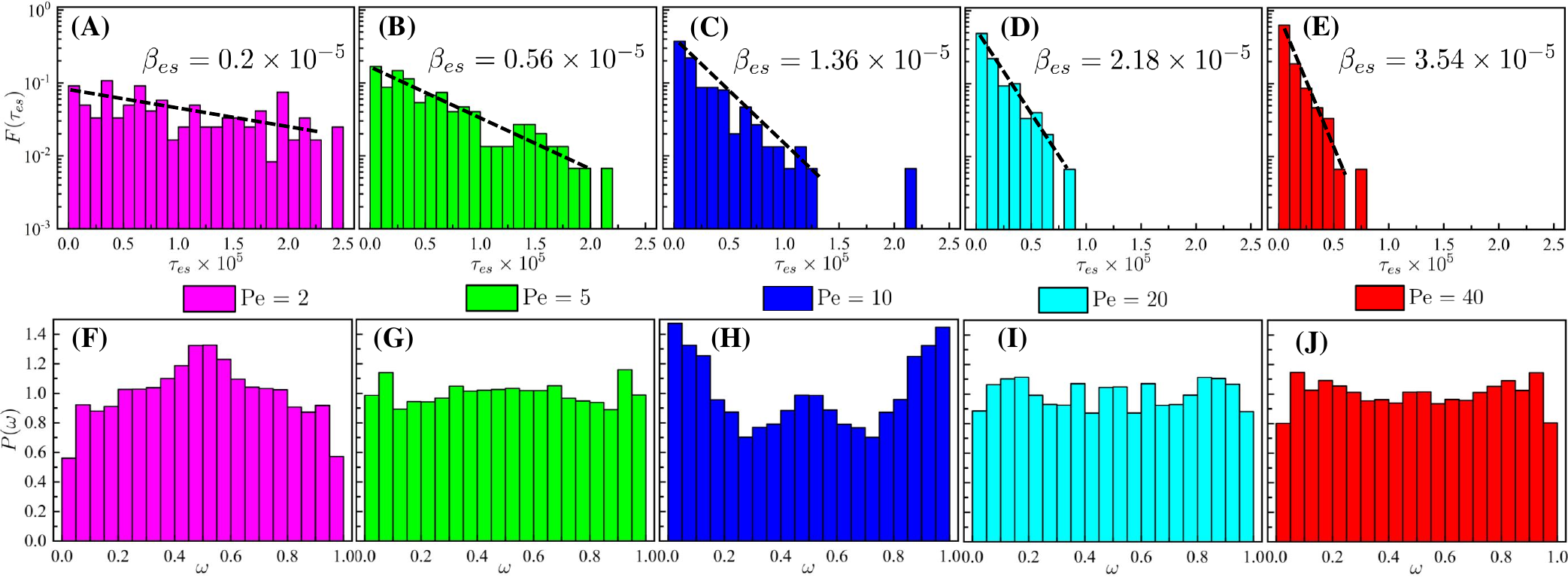}
    \caption{Plots of the first escape times distributions $F(\tau_{es})$ and corresponding uniformity index distributions $P(\omega)$ of self-driven nanorod are shown in (A-E) and (F-J), respectively, at Pe $= 2, 5, 10, 20, 40$, for confinement with a single window. $P(\omega)$ has a unimodal behavior at the lowest values of Pe $= 2$, but not for higher values of Pe. One can see that a significant fraction of large-time data is beyond short-time exponential fittings.}
    \label{fig:Dist_FPT_UI}
\end{figure*}

\noindent Subsequently, we depict the coefficient of variation $CV$ (ratio of the standard deviation to the mean) in Fig.~\ref{fig:MFPT_CV}(B), which measures the fluctuations in the first escape times \cite{mattos2012first,co2017stochastic,biswas2020first}.

\begin{equation}
    CV = \dfrac{\sqrt{\left < \tau_{es}^2 \right > - \left < \tau_{es} \right >^2}}{\left < \tau_{es} \right >}
\end{equation}

\noindent where $\left < \tau_{es} \right >$ and $\left < \tau_{es}^2 \right >$ are the first and second moments of the first escape times of a self-propelled nanorod. It is also known as ``relative standard deviation". A low $CV$ value indicates little variation in first escape times, whereas a high $CV$ indicates more variation. For confinement with single and three opening(s), we see an interesting non-monotonic behavior of $CV$ as a function of Pe (Fig.~\ref{fig:MFPT_CV}(B)). When considering confinement with a single opening, the coefficient of variation ($CV$) exhibits a pattern of variation. Initially, the $CV$ increases from Pe $= 2$ to $10$ and reaches its maximum at Pe $= 10$, then decreases as Pe is further increased up to 20, then it has a small rise when Pe is increased from 20 to 40. Similarly, the $CV$ exhibits non-monotonic behavior for a domain with three openings. In contrast, $CV$ always increases with Pe in the case of five openings. However, $CV$ is just a number and it is unclear how to anticipate it. In general, when $CV$ exceeds unity, implying that the standard deviation (numerator) is greater than the mean (denominator), the mean value $\left < \tau_{es} \right >$ does not adequately represent the stochastic time scales \cite{mattos2012first}. Therefore, we must look at the full distribution $F(\tau_{es})$ of the active nanorod's first escape times $\tau_{es}$ [Fig.~\ref{fig:Dist_FPT_UI}(A-E)]. We also analyze the corresponding distribution $P(\omega)$ of the uniformity index $\omega$ [Fig.~\ref{fig:Dist_FPT_UI}(F-J)], a method to quantify the trajectory-to-trajectory fluctuations in time scales \cite{mattos2012first,biswas2020first,mattos2014trajectory}, which is defined as follows,

\begin{equation}
    \omega = \dfrac{\tau'_{es}}{\tau'_{es} + \tau''_{es}}
\end{equation}

\noindent where $\tau'_{es}$ and $\tau''_{es}$ are two random first escape times for an active nanorod. It is clear from the above expression of $\omega$, if $\tau'_{es} \gg \tau''_{es}$, $\omega \approx \dfrac{\tau'_{es}}{\tau'_{es}} \approx 1$. On the other hand, if $\tau'_{es} \ll \tau''_{es}$, $\omega \approx \dfrac{\tau'_{es}}{\tau''_{es}} \approx 0$. Interestingly, if $\tau'_{es} \approx \tau''_{es}$, $\omega \approx \dfrac{1}{2}$. Therefore by definition, $\omega$ varies from 0 to 1 and $\omega \approx \dfrac{1}{2}$, when the trajectory-to-trajectory fluctuation is minimal. In contrast, distribution $P(\omega)$ is expected to become broader and develop peaks near 0 and 1 for large fluctuations in the values of $\tau_{es}$. \\

\noindent We plot $F(\tau_{es})$ and $P(\omega)$ of the first escape dynamics in Fig.~\ref{fig:Dist_FPT_UI} for various activities (Pe). Firstly, we focus on the first escape times distribution $F(\tau_{es})$ and $P(\omega)$ for the case with a single opening. For Pe $= 2$, $F(\tau_{es})$ has a broader distribution over a time scale from around $5\times 10^4$ to $2\times 10^5$, where each of the trajectories has a nearly equal probability of occurring [Fig.~\ref{fig:Dist_FPT_UI}(A)]. However, the distribution becomes narrower, shifts towards shorter times, and shows good exponential fittings with a larger value of $\beta_{es}$ as Pe increases [Fig.~\ref{fig:Dist_FPT_UI}(B-E)]. We also observe a sharper drop of $F(\tau_{es})$ with increasing Pe, which can clearly be seen from increasing value of $\beta_{es}$ with Pe [Fig.~\ref{fig:Dist_FPT_UI}(B-E)]. We see that at Pe = $5$, $F(\tau_{es})$ has a typical decay over $\sim 2\times 10^3$, but there is a sizable portion of large time data ($\tau_{es} \sim 2\times 10^5$), which is beyond exponential fitting, as shown in Fig.~\ref{fig:Dist_FPT_UI}(B). This implies that a significant fraction of the observed trajectories is short in addition to the long trajectories. We understand that most of the time, the active nanorod quickly moves toward the boundary from the center, finds a small pore, and escapes through it [see ESI, Movie S10]. On the other hand, for the same Pe, the nanorod moves along the boundary repeatedly until it gets a proper alignment towards a small opening and escapes through it [see ESI, Movie S11]. This is more prominent for Pe = $10$, where $F(\tau_{es})$ has an initial exponential decay over a short typical time $\sim 10^3$, implying that a fraction of the observed trajectories is short, which points out that the nanorod quickly escapes [Fig.~\ref{fig:Dist_FPT_UI}(C)]. One such trajectory is shown in Fig.~S7(A) in ESI. Yet, in the same figure, we notice that a significant fraction of time scales are much larger ($\tau_{es} \sim2\times 10^5$), which is beyond exponential fitting. This signifies that the active nanorod moves along the boundary repeatedly for an extended period of time before exiting from confinement. Such a trajectory may be seen in Fig.~S7(B) in ESI. It is also reflected in the high value of $CV$ for confinement with a single opening, $n_o = 1$ [Fig.~\ref{fig:MFPT_CV}(B)]. For the nanorod with higher self-propulsion (\textit{e.g.,} Pe = 20, 40), $F(\tau_{es})$ displays narrower distributions with initial exponential decay over a short shorter time ($\sim 5\times 10^2$). These fluctuations in the time scales are further supported by the behavior of $P(\omega)$. \\

\noindent We find that the shape of the distributions is highly dependent on Pe, as shown in Fig.~\ref{fig:Dist_FPT_UI}(F-J) for confinement with a single opening. For Pe $ = 2$, $P(\omega)$ represents the unimodal, bell-shaped distribution where the peak is at around $\omega = \frac{1}{2}$, indicating that most pairs of first escape times are similar and the trajectory-to-trajectory fluctuations of first escape time are not prominent. Conversely, the shape of distributions $P(\omega)$ changes on varying the activity. One can notice that $P(\omega)$ exhibits a broader distribution for Pe $ = 5$. For Pe $ = 10$, $P(\omega)$ shows bimodal distribution with a local minimum at around $\omega = \dfrac{1}{2}$ and two maxima close to $\omega = 0$ and $\omega = 1$ [Fig.~\ref{fig:Dist_FPT_UI}(H)]. As discussed above, $CV$ also has a maximum value at Pe $= 10$ [Fig.~\ref{fig:MFPT_CV}(B)] in the case of a single opening. This may also be seen from the snapshots of the trajectories in Fig.~S7(A-B) in ESI. Moreover, the shape of $P(\omega)$ becomes increasingly broader from Fig.~\ref{fig:Dist_FPT_UI}(H) to Fig.~\ref{fig:Dist_FPT_UI}(J) on increasing Pe from 10 to 40. This implies, in these cases, the MFET cannot be considered an adequate measure of the actual behavior. Similar distributions have been observed earlier in experimental and theoretical studies of particles inside various confined geometries \cite{mattos2012first,biswas2020first,mattos2014trajectory,biswas2023}. It is important to note that the specific behavior of the fluctuations in the first escape times depends on the details of the system, the nature of self-propulsion, and the characteristics of the confinement. On the other hand, the first escape statistics exhibit large trajectory-to-trajectory fluctuation in the case of multiple openings, which is reflected in the high values of $CV$, the presence of multiple timescales in $F(\tau_{es})$, and multimodality in $P(\omega)$ [Fig.~S8 and Fig.~S9, ESI], and it is maximum for confinement with three openings at higher Pe (Fig.~\ref{fig:MFPT_CV}(B)). In Fig.~S7(C, E) in ESI, we display short trajectories, where the nanorod quickly reaches close to the boundary and escapes for a given Pe and $n_o$. Whereas for the same Pe and $n_o$, the active nanorod moves along the circular confinement repeatedly, which results in long circular trajectories, as shown in Fig.~S7(D, F) in ESI. In these cases, the MFET is not representative of the actual behavior. The presence of multiple small windows may introduce spatial heterogeneity and diverse exit pathways, which can lead to larger fluctuations in the first exit times of the active nanorod. This may be seen clearly in ESI videos, namely Movies S5, S6, S8, S9, and S12-S15, for the confinements with multiple openings. \\

\begin{figure*}[t]
    \centering
    \includegraphics[width=0.99\linewidth]{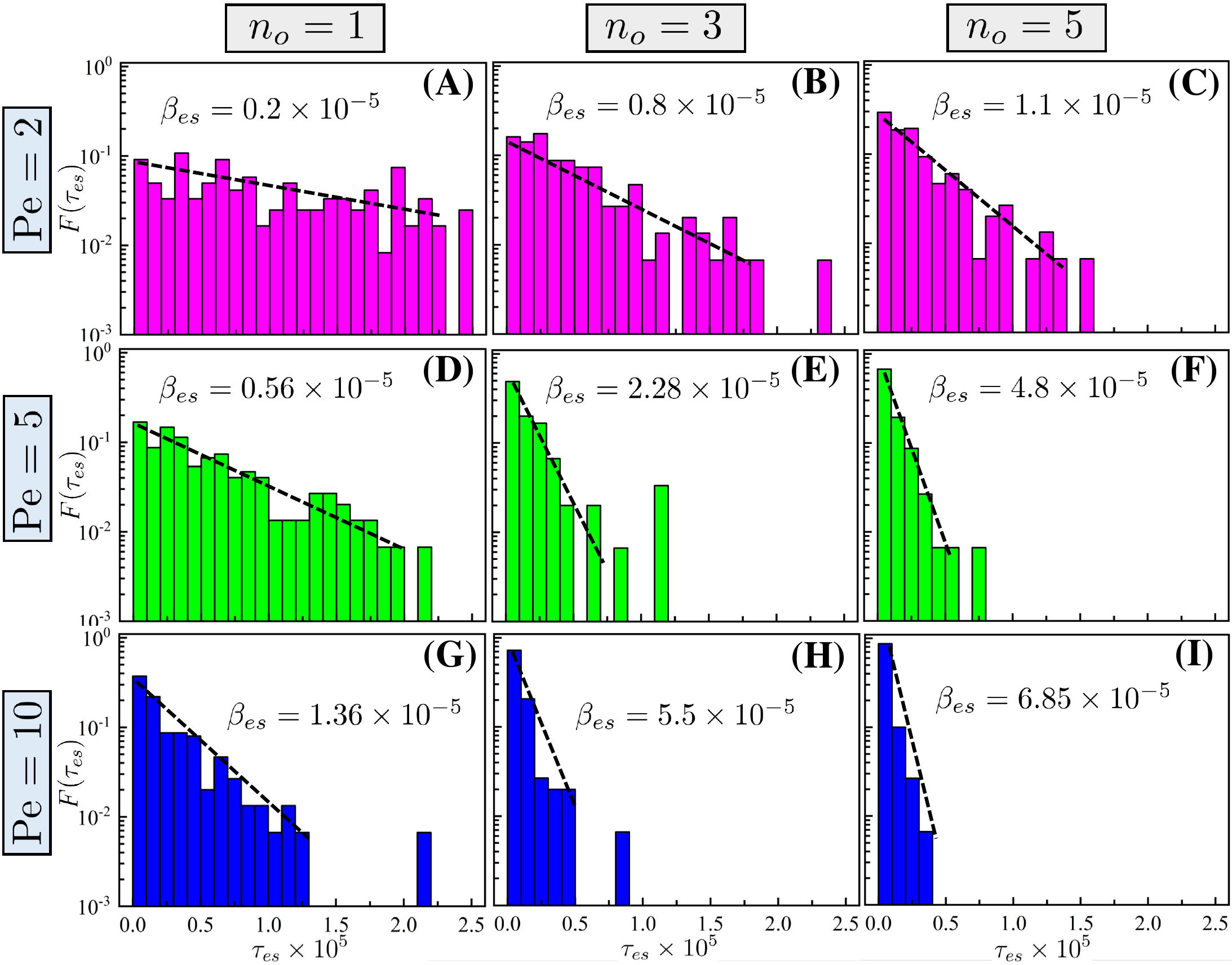}
    \caption{A comparison of $F(\tau_{es})$ for different values of Pe (from top to bottom for a given value of $n_{o}$) and the number of openings $n_o$ (from left to right for a given value of Pe). We see a significant fraction of $\tau_{es}$ shifts towards shorter times on increasing Pe and $n_o$. Short to moderate time data show good exponential fittings and the slope $\beta_{es}$ increases with activity (Pe) and $n_o$.}
    \label{fig:Dist_FPT_Comp}
\end{figure*}

\noindent In addition, we compare the distributions $F(\tau_{es})$ of an active nanorod inside the confinement by varying the numbers of small openings as a function of self-propulsion [Fig.~\ref{fig:Dist_FPT_Comp}]. This comparative analysis enables us to understand the relationship between the number of openings, self-propulsion, and the resulting first escape times distribution for the active nanorod. From the analysis of Fig.~\ref{fig:Dist_FPT_Comp}, we notice a consistent trend where the distributions shift towards shorter time scales with larger values of $\beta_{es}$ on increasing the number of openings for a fixed value of Pe, indicating, on average, the active nanorod takes less time to escape the confinement. As shown above in Fig.~\ref{fig:MFPT_CV}(B), the trajectory-to-trajectory fluctuations are more prominent for the confinement with multiple openings, as further shown in Fig.~S8 and Fig.~S9 in ESI. This observation suggests that the number of openings also plays a significant role in the nanorod's escape dynamics.  \\

\section{Conclusion}\label{sec:Conclusion}
\noindent Motivated by a growing interest in controlling the transport and diffusion of microswimmers in confined domains, we performed computer experiments to investigate the escape dynamics of a self-propelled nanorod from the circular confinements with single and multiple opening(s) through which the nanorod can escape. We repeated the simulations to obtain statistically significant information on the time taken by the nanorod to find a narrow opening and successfully exit through this. First, we simulated the passive nanorod and observed that the nanorod moves inside the circular confinement. On the other hand, our simulation results displayed that the self-driven nanorod escapes through a narrow opening due to the persistent motion if it has the orientation towards the opening and enters again through the gateway as it gets a chance. The frequency of coming out of and getting into the confinement increases with the activity as well as the number of openings on the boundary that reflects in radial probability distribution plots. For the passive case (Pe $= 0$), the probability of finding the particle inside the confinement is uniformly distributed. In contrast, in the case of the active nanorod, the probability distributions in each of the cases exhibit two peaks close to the boundary due to the faster and more persistent motion of the self-propelled nanorod. Interestingly, the radial probability density profiles become narrower with self-propulsion due to the impermeable boundary, and peaks shift more towards the confinement wall. For a deeper understanding of the escape dynamics of the active nanorod, we analyzed the mean first escape time, coefficient of variation, and uniformity index distributions. Interestingly, it has been found that the mean escape time reduces as the self-propulsion increases, as would be intuitively expected, while their fluctuations vary non-monotonically. We also observed a non-monotonic behavior by changing the number of openings within the system. We hope that our current study will help in designing self-driven nanorobots assigned to navigate through narrow channels \cite{medina2018micro,park2017multifunctional,mujtaba2021micro,mitchell2021engineering}.

\section*{Acknowledgements}
\noindent P. K. thanks IIT Bombay for Institute Postdoctoral Fellowship. R. C. acknowledges Science and Engineering Research Board (SERB) for funding (Project No. MTR/2020/000230 under MATRICS scheme) for funding. We acknowledge the SpaceTime-2 supercomputing facility at IIT Bombay for the computing time. P. K. would like to thank Dr. Koushik Goswami and Rajiblochan Sahoo for critically reading the manuscript and discussing it. \\ \\


\noindent \textbf{DATA AVAILABILITY} \\

\noindent The data that supports the findings of this study are available within the article [and its supplementary material]. \\

\renewcommand{\thefigure}{S\arabic{figure}}
\setcounter{figure}{0}
\setcounter{equation}{0}
\renewcommand{\theequation}{\Roman{equation}}
\appendix

\noindent \textbf{SUPPLEMENTARY MATERIAL} \\

\begin{figure*}[ht]
    \includegraphics[width=0.6\linewidth]{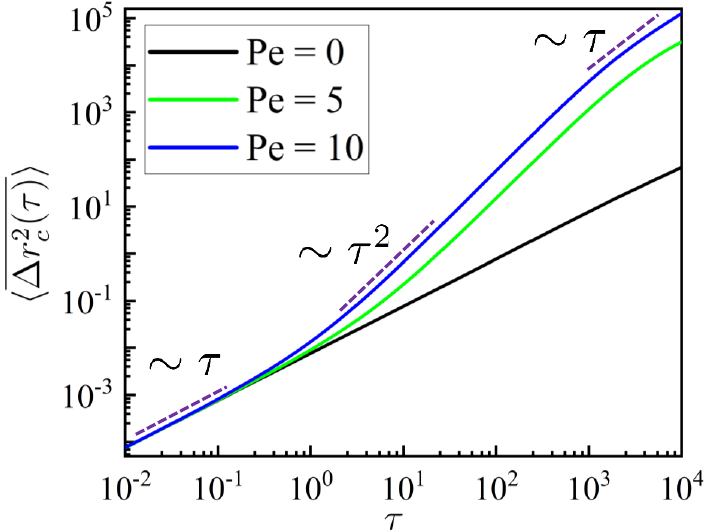}
    \caption{Log-log plot of  $\left<\overline{\Delta r_\text{c}^{2}(\tau)}\right>$ for the self-propelled nanorod at different Pe.}
    \label{fig:MSD}
\end{figure*}

\noindent The method and parameter validation is carried out for a nanorod in free space. The translational $\left(\left<\overline{\Delta r_\text{c}^{2}(\tau)}\right>\right)$ mean square displacement is calculated. From the plot (Fig.~\ref{fig:MSD}) for Pe = 0, we have computed the thermal translational diffusion coefficient, $D_{T} = 1.87 \times 10^{-5}$ and friction coefficient, $\gamma = 5.33 \times 10^4$. \\

\noindent \textbf{Movies} \\
\noindent Movie\_S1: Brownian dynamics simulation of the passive (Pe = 0) nanorod for the confinement with a single narrow opening ($n_o = 1$).

\noindent Movie\_S2: Brownian dynamics simulation of the passive (Pe = 0) nanorod for the confinement with three narrow openings ($n_o = 3$).

\noindent Movie\_S3: Brownian dynamics simulation of the passive (Pe = 0) nanorod for the confinement with five narrow openings ($n_o = 5$).

\noindent Movie\_S4: Brownian dynamics simulation of the active (Pe = 5) nanorod for the confinement with a single narrow opening ($n_o = 1$).

\noindent Movie\_S5: Brownian dynamics simulation of the active (Pe = 5) nanorod for the confinement with three narrow openings ($n_o = 3$).

\noindent Movie\_S6: Brownian dynamics simulation of the active (Pe = 5) nanorod for the confinement with five narrow openings ($n_o = 5$).

\noindent Movie\_S7: Brownian dynamics simulation of the active (Pe = 20) nanorod for the confinement with a single narrow opening ($n_o = 1$).

\noindent Movie\_S8: Brownian dynamics simulation of the active (Pe = 20) nanorod for the confinement with three narrow openings ($n_o = 3$).

\noindent Movie\_S9: Brownian dynamics simulation of the active (Pe = 20) nanorod for the confinement with five narrow openings ($n_o = 5$).

\noindent Movie\_S10: Brownian dynamics simulation of the active (Pe = 10) nanorod for the confinement with a single narrow opening ($n_o = 1$). Here, the active nanorod quickly moves toward the boundary from the center, finds a small pore, and escapes through it.

\noindent Movie\_S11: Brownian dynamics simulation of the active (Pe = 10) nanorod for the confinement with a single narrow opening ($n_o = 1$). The active nanorod moves along the boundary repeatedly for a long time.

\noindent Movie\_S12: Brownian dynamics simulation of the active (Pe = 5) nanorod for the confinement with three narrow openings ($n_o = 3$). Here, the nanorod has a small trajectory.

\noindent Movie\_S13: Brownian dynamics simulation of the active (Pe = 5) nanorod for the confinement with three narrow openings ($n_o = 3$). Here, the active nanorod moves along the boundary repeatedly for a long time, it has a long circular trajectory.

\noindent Movie\_S14: Brownian dynamics simulation of the active (Pe = 40) nanorod for the confinement with five narrow openings ($n_o = 5$). Here, the active nanorod quickly escapes through a narrow opening.

\noindent Movie\_S15: Brownian dynamics simulation of the active (Pe = 40) nanorod for the confinement with five narrow openings ($n_o = 5$). The active nanorod moves along the boundary repeatedly for a long time to escape from the confinement.

\noindent For the passive case (Pe = 0), $\left<\overline{\Delta r_{c}^{2}(\tau)}\right>$ is always diffusive $\left(\left<\overline{\Delta r_{c}^{2}(\tau)}\right> \sim \tau\right)$ with the diffusion coefficient $D_{T}$. In case of the self-propelled nanorod, $\left<\overline{\Delta r_{c}^{2}(\tau)}\right>$ exhibits three distinct regions: diffusive at short time, superdiffusive region at the intermediate time which scales as $\left<\overline{\Delta r_{c}^{2}(\tau)}\right>$ $\sim \tau^2$. At longer time, $\left<\overline{\Delta r_{c}^{2}(\tau)}\right>$ becomes linear in time with an enhanced diffusion coefficient.

\begin{figure*}[h!]
    \centering
    \includegraphics[width=0.99\linewidth]{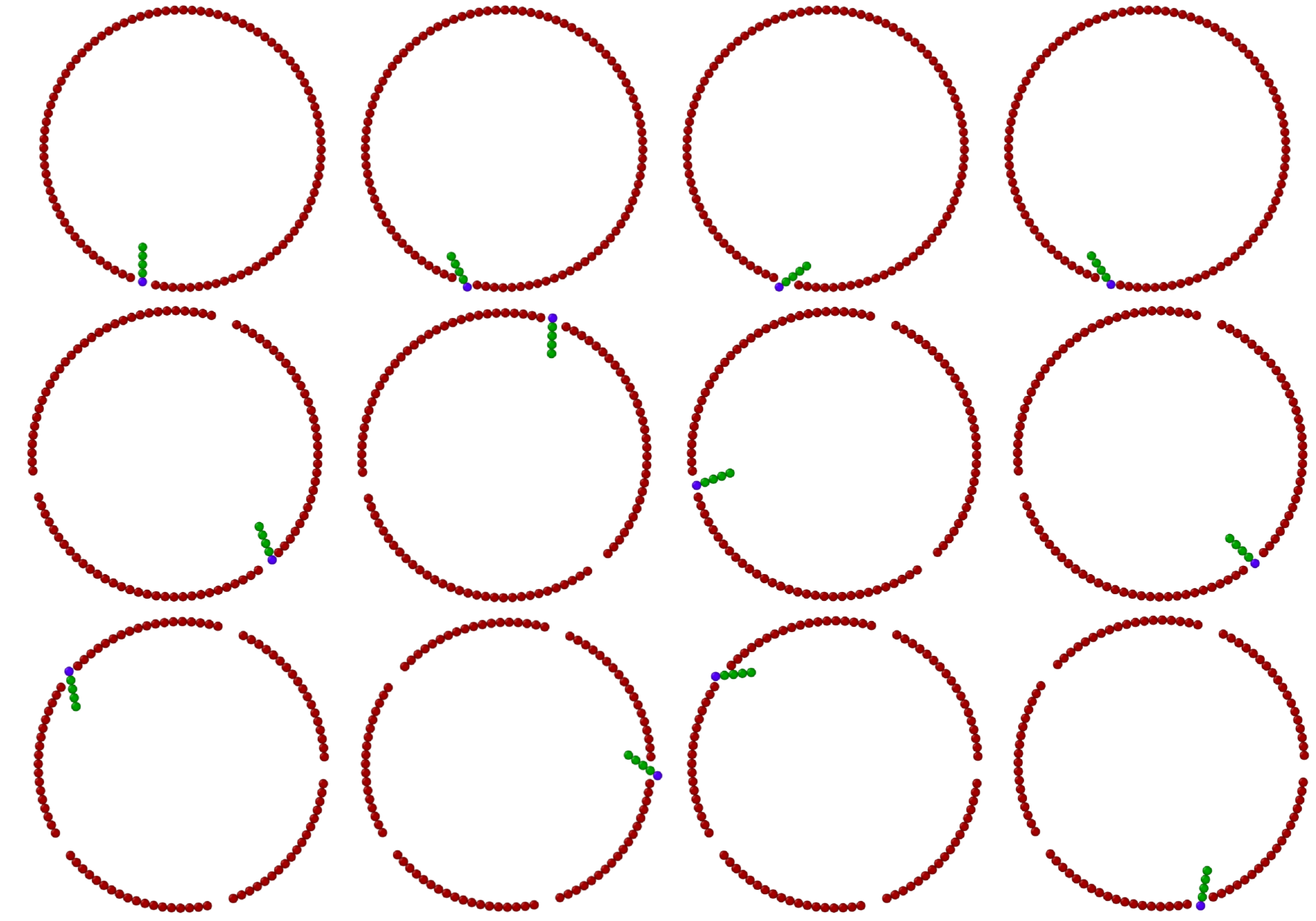}
    \caption{Here are some snapshots from the simulations where the nanorod successfully escapes through the narrow opening(s) with different orientations.}
    \label{fig:Orientation_escape}
\end{figure*}

\begin{figure*}[h!]
    \centering
    \includegraphics[width=0.99\linewidth]{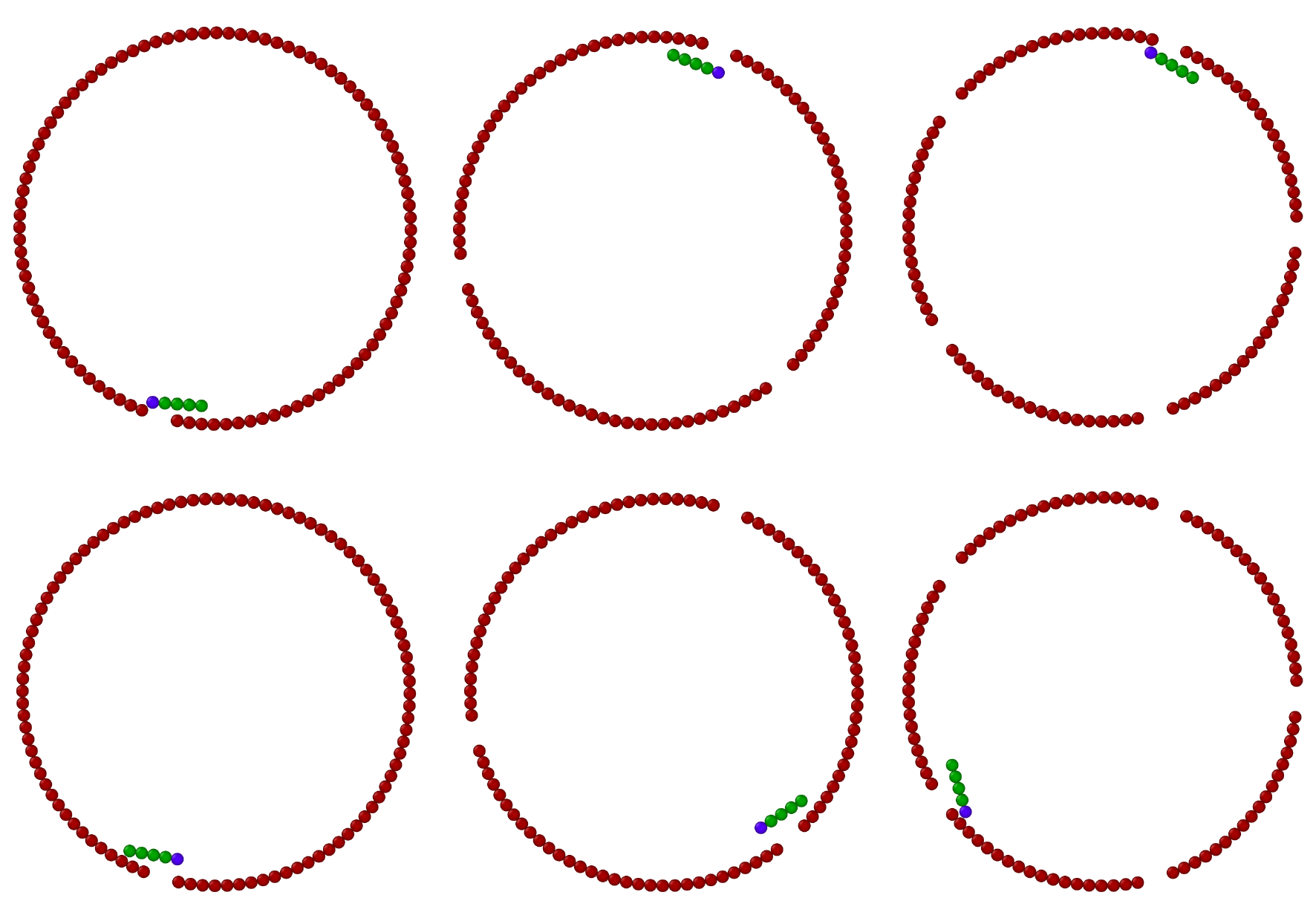}
    \caption{Here are some snapshots from the simulations depicting instances where the nanorod is unable to escape through the narrow opening(s) due to its improper orientation as it remains nearly parallel to the wall.}
    \label{fig:Orientation_no-escape}
\end{figure*}

\begin{figure*}[ht]
    \includegraphics[width=0.9\linewidth]{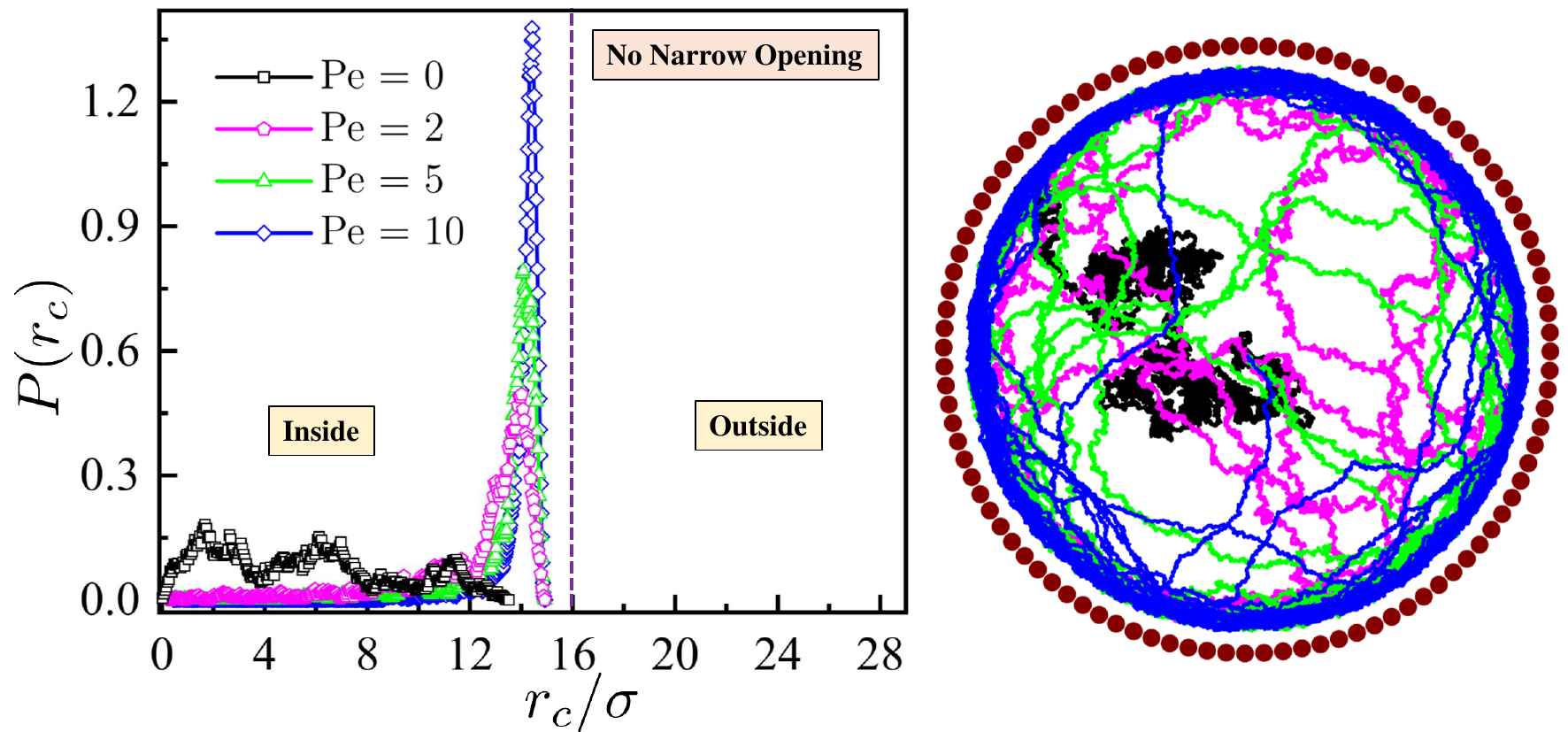}
    \caption{The radial probability distributions $P(r_{c})$ of the nanorod for the confinement without opening, $n_{o} = 0$ (left) and  the corresponding COM trajectories  (right) for different Pe.}
    \label{fig:RPDF_N_0}
\end{figure*}

\begin{figure*}[ht]
    \includegraphics[width=0.95\linewidth]{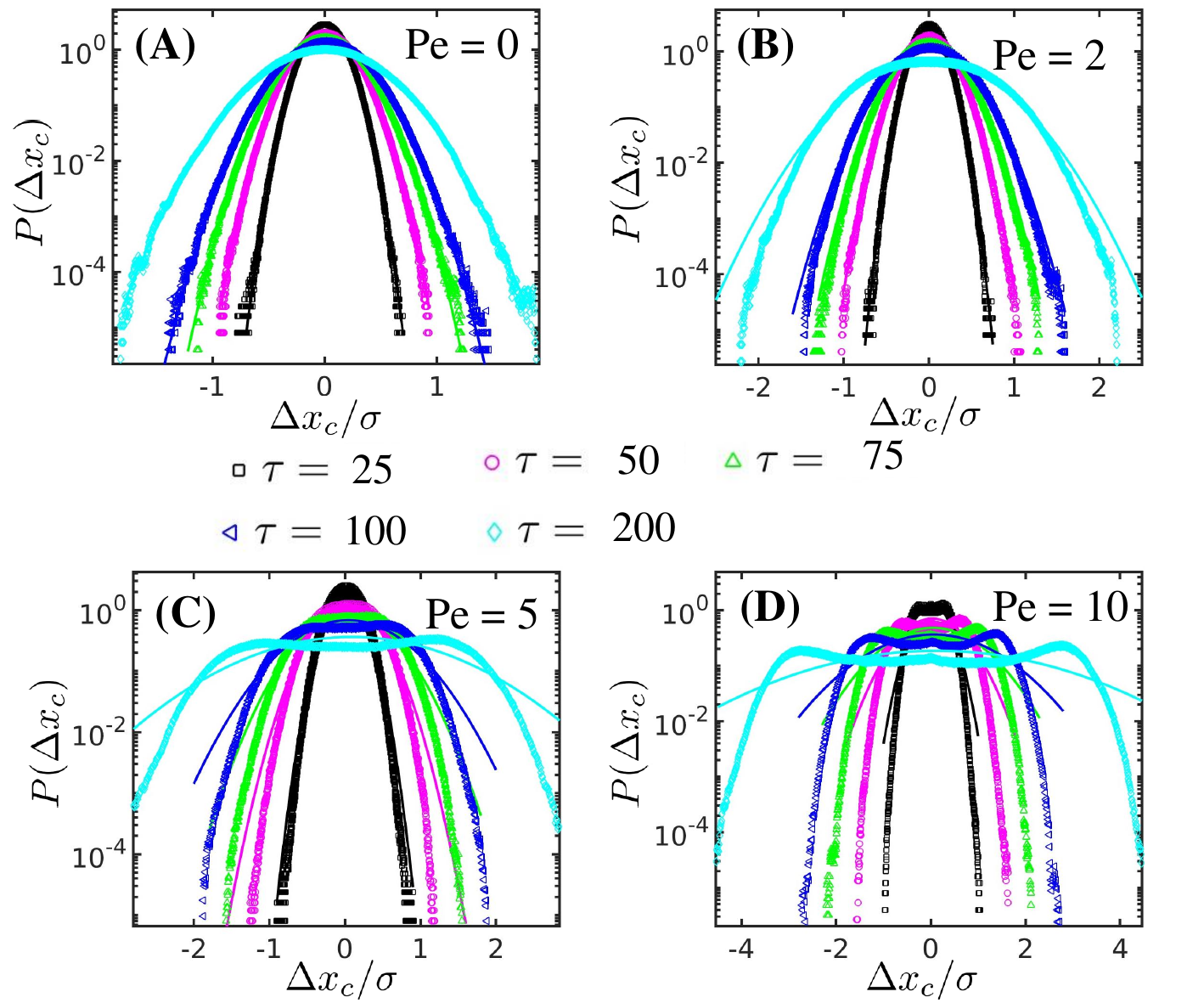}
    \caption{Plots (A)-(D) represent displacement distribution functions $P(\Delta x_{c})$ of a nanorod at different lag times $\tau = 25, 50, 75, 100, 200$ for the confinement with three narrow openings. Solid lines represent the Gaussian fittings.}
    \label{fig:FEST_UI_3O}
\end{figure*}

\begin{figure*}[h!]
    \centering
    \includegraphics[width=0.99\linewidth]{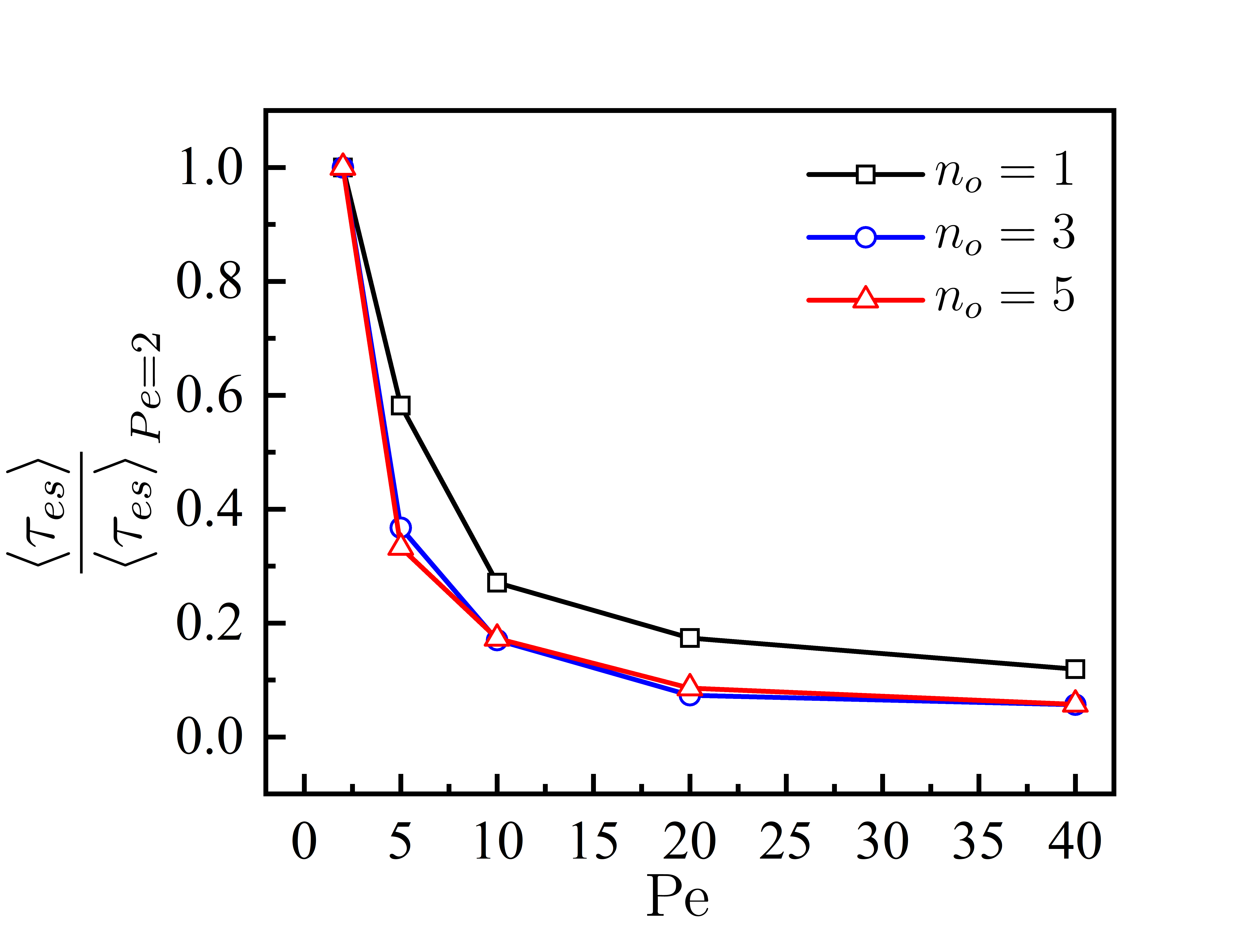}
    \caption{Plot of $\dfrac{\left < \tau_{es} \right >}{\left < \tau_{es} \right >_{Pe = 2}}$ \textit{versus} Pe.}
    \label{fig:MFPT_comp}
\end{figure*}

\begin{figure*}[ht]
    \includegraphics[width=0.95\linewidth]{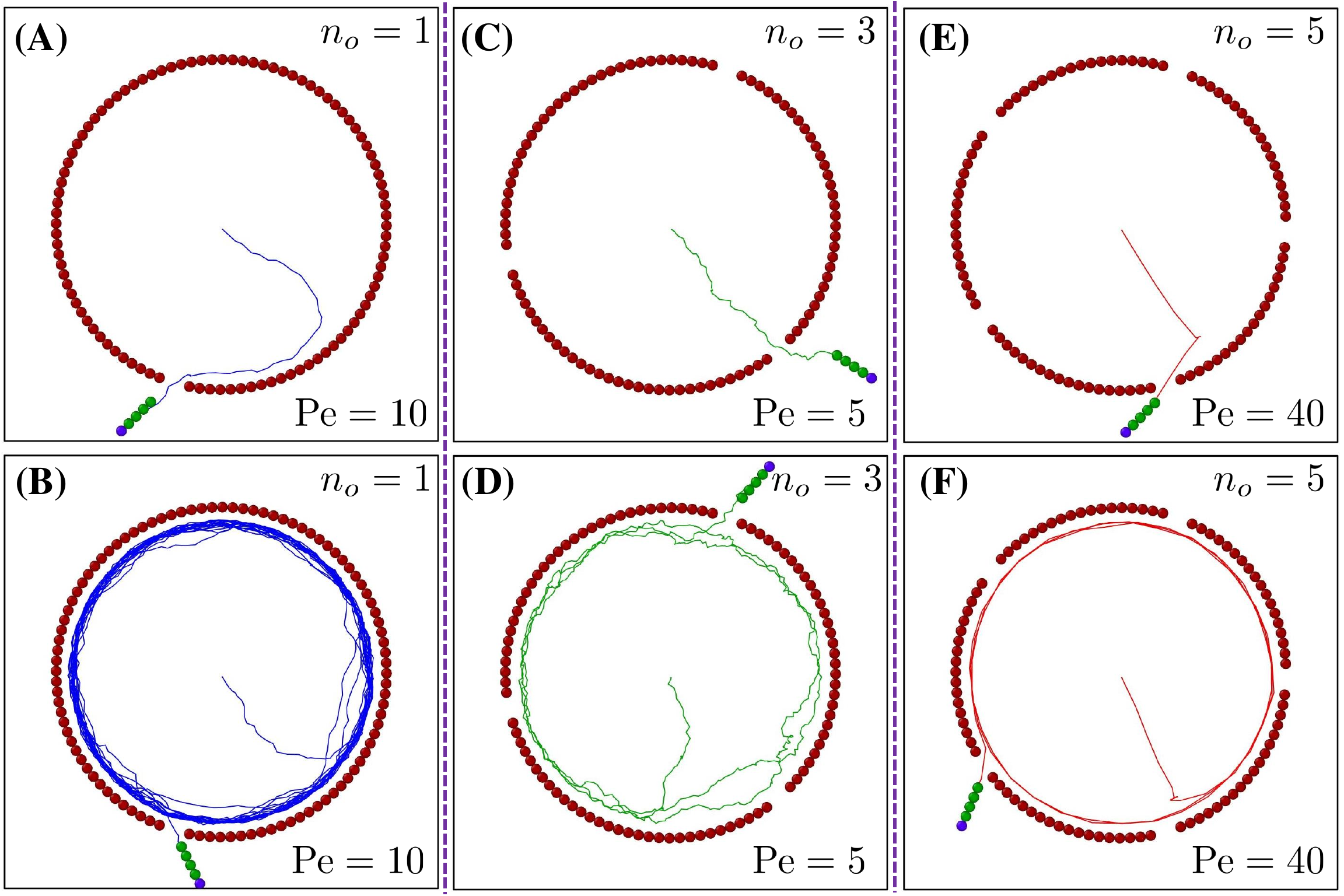}
    \caption{The COM trajectories of the nanorod in (A) short time escape trajectory, (B) long time escape trajectory at Pe = 10 for a single opening case, (C) short time escape trajectory, (D) long time escape trajectory at Pe = 5 for three openings case, (E) short time escape trajectory, and (F) long time escape trajectory at Pe = 40 for five openings case}
    \label{fig:Movie_traj}
\end{figure*}

\begin{figure*}[ht]
    \includegraphics[width=0.95\linewidth]{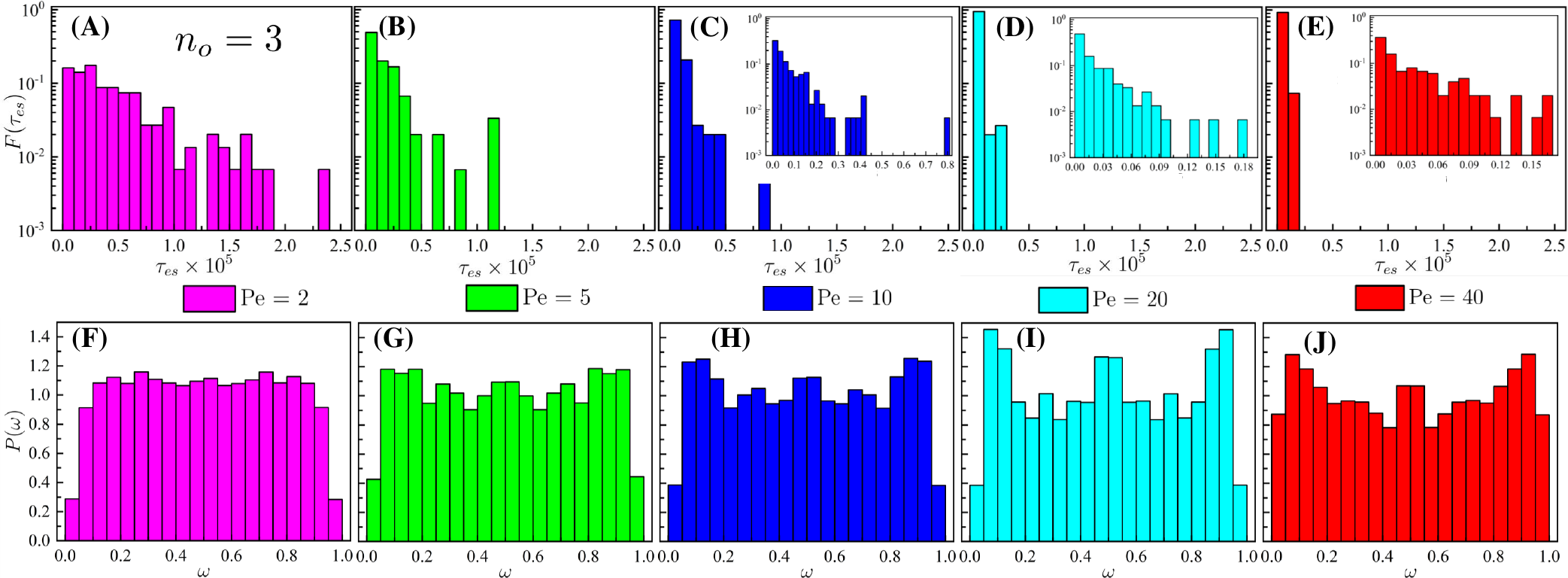}
    \caption{Plots of the first escape times distributions $F(\tau_{es})$ and corresponding uniformity index distributions $P(\omega)$ of self-driven nanorod are shown in (A-E) and (F-J), respectively, at Pe $= 2, 5, 10, 20, 40$, for confinement with three openings.}
    \label{fig:FEST_UI_3O}
\end{figure*}

\begin{figure*}[ht]
    \includegraphics[width=0.95\linewidth]{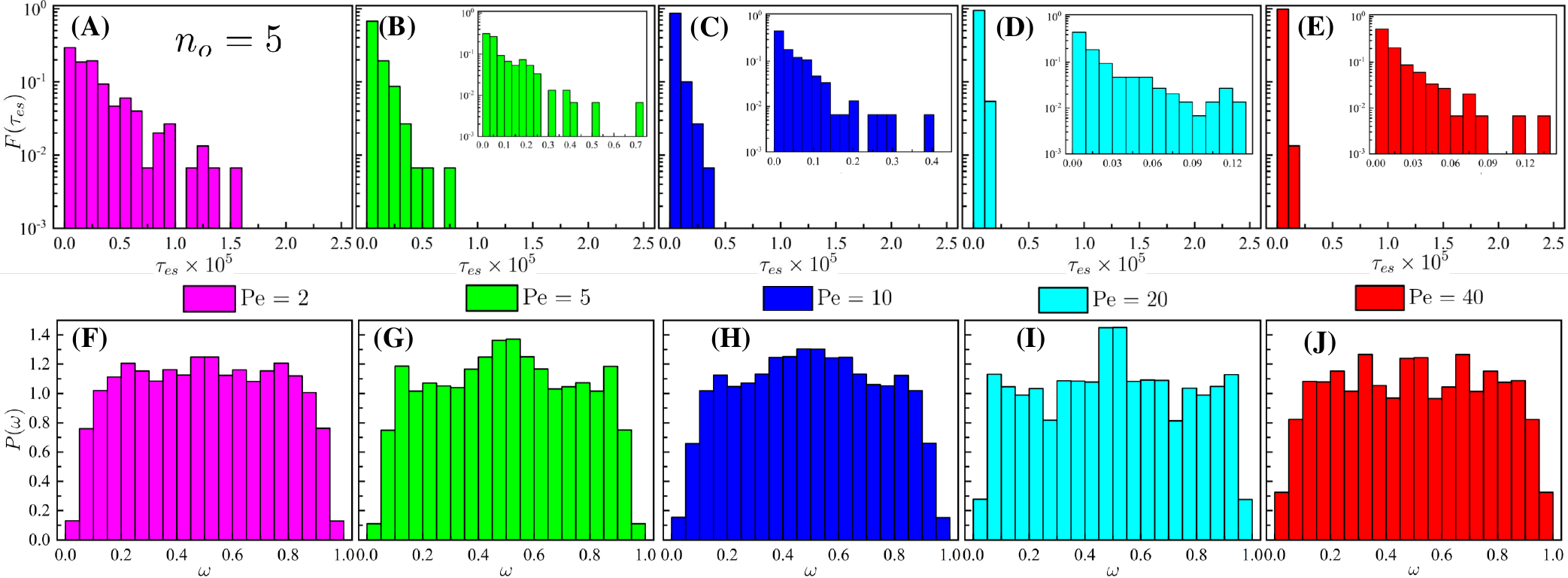}
    \caption{Plots of the first escape times distributions $F(\tau_{es})$ and corresponding uniformity index distributions $P(\omega)$ of self-driven nanorod are shown in (A-E) and (F-J), respectively, at Pe $= 2, 5, 10, 20, 40$, for confinement with five openings.}
    \label{fig:FEST_UI_5O}
\end{figure*}

\clearpage
\newpage

%

\end{document}